\documentclass{svjour3}                     % onecolumn (standard format)
\smartqed  % flush right qed marks, e.g. at end of proof
\usepackage{graphicx}
\begin{document}

\title{Few-body decay and recombination in nuclear astrophysics
\thanks{Presented at the 21st European Conference on Few-Body Problems in
Physics, Salamanca, Spain, 30 August - 3 September 2010}}

%\title{Recombination and three-body decays of light nuclei in stellar 
%environments % Insert your title here%\Thanks{Grants or other notes

\titlerunning{Recombination and three-body decays}        
% if too long for running head

\author{A.S. Jensen \and D.V. Fedorov \and 
R. de Diego \and E. Garrido \and R. \'Alvarez-Rodr\'{\i}guez }


\institute{A.S. Jensen and D.V. Fedorov \at Department of Physics and 
Astronomy,  Aarhus University, DK-8000 Aarhus C, Denmark \\
              Tel.: +45-8942 3655 and +45-8942 3651 \\              
\email{asj@phys.au.dk and fedorov@phys.au.dk}           %  \\
\and R. de Diego and E. Garrido 
\at Instituto de Estructura de la Materia, CSIC, Serrano 123, E-28006
Madrid, Spain 
\and R. \'Alvarez-Rodr\'{\i}guez 
\at Departamento de F\'{\i}sica At\'omica, Molecular y Nuclear, 
Universidad Complutense de Madrid, E-28040 Madrid, Spain }

\date{Received: date / Accepted: date}

\maketitle

\begin{abstract}
  Three-body continuum problems are investigated for light nuclei of
  astrophysical relevance. We focus on three-body decays of resonances
  or recombination via resonances or the continuum background. The
  concepts of widths, decay mechanisms and dynamic evolution are
  discussed. We also discuss results for the triple $\alpha$ decay in
  connection with $2^+$ resonances and density and temperature
  dependence rates of recombination into light nuclei from
  $\alpha$-particles and neutrons.  
\keywords{Few-body problems \and Astrophysics \and Unstable nuclei}
\end{abstract}

\section{Introduction}
\label{intro}

Few-body problems, astrophysics and unstable nuclei are naturally
linked through light nuclei being formed in few-body recombination
reactions in astrophysical environments.  The inverse reactions of
few-body decays of light nuclei are equivalent through detailed
balance. The state of the art in few-body physics is that all two-body
problems and essentially all bound state three-body problems are fully
solved. The next in line is three-body continuum problems where lots
of information presently accumulate from kinematically complete and
accurate measurements of nuclear decay processes.  The simplest of
these examples are nuclear three-body decays of a (many-body)
resonance.  A large variety of decay mechanisms is possible due to the
continuous distribution of energy and momentum between the three
particles in the final state. We shall here focus on genuine
three-body continuum computations. We shall especially emphasize the
new concepts employed or discovered, and illustrate with examples of
relevance in nuclear astrophysics.

\section{Definitions}
\label{sec:0}

The minimum needed for the discussions is the hyperradius $\rho$
defined as the mass weighted mean square radius coordinate:
\begin{equation} \label{e20}
 m \rho^2 =  \frac{1}{m_1+m_2+m_3} \sum_{k} m_k ({\bf r}_k -{\bf R})^2  \;\; 
\end{equation}
where $m$ is an arbitrary mass, $m_k$ and $\bf r_k$ are mass and
position of the $k$'th particle and $\bf R$ is the centre-of-mass
coordinates. The total wave function is $\psi =\rho^{-5/2} \sum_n
  f_n(\rho) \phi_n(\rho,\Omega)$, where $\Omega$ denotes the five
 angular coordinates.  The differential equations for one radial function 
$f_n$ are \cite{nie01}
\begin{eqnarray} \label{e50}
\left[ -\frac{d^2}{d\rho^2} + \frac{\lambda(\rho)+ 15/4}{\rho^2} + Q(\rho) -
\frac{2m E }{\hbar^2} \right] f_n(\rho) = couplings \; ,\; 
 Q(\rho)  = \langle \phi | \frac{\partial^2}{\partial \rho^2} | \phi
  \rangle_{\Omega} \; ,
\end{eqnarray}
where $E$ is the energy, $\lambda$ is obtained from the angular
equations, and $Q$ is the non-adiabatic diagonal coupling strongly correlated to the
variation of the angular wave function. 

\begin{figure}
\vspace*{-0.5cm}
  \includegraphics[width=0.75\textwidth,angle=-90]{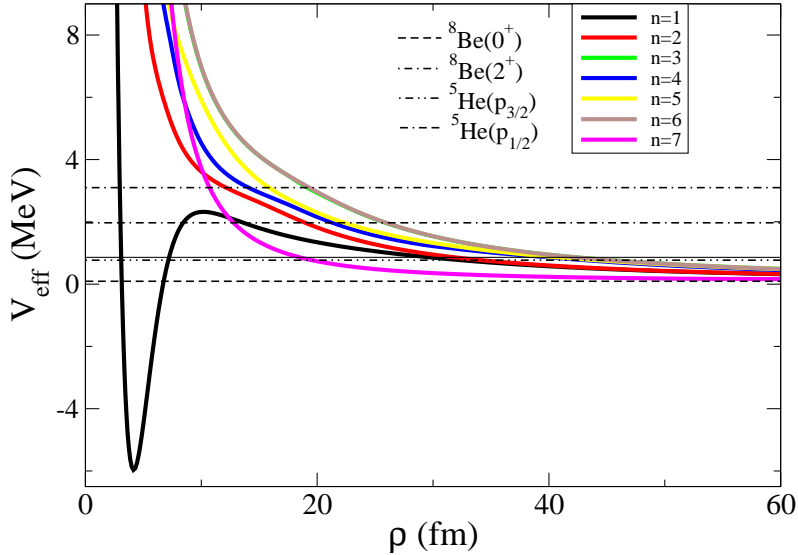}
%/usr/users/asj/raq/be9fig1.ps}
%  \includegraphics{example.eps}
\vspace*{-0.5cm}
\caption{The real parts for $\theta = 0.1$ of the lowest adiabatic
  potentials, including the three-body potential, for the $0.856$~MeV
  $^{9}$Be($5/2^{-}$)-resonance (horizontal full line) as function of
  $\rho$. }
\label{fig:1}       % Give a unique label
\end{figure}

\section{Concepts in three-body decays}
\label{sec:1}

\begin{figure}
\vspace*{-0.5cm}
  \includegraphics[width=0.75\textwidth,angle=-90]{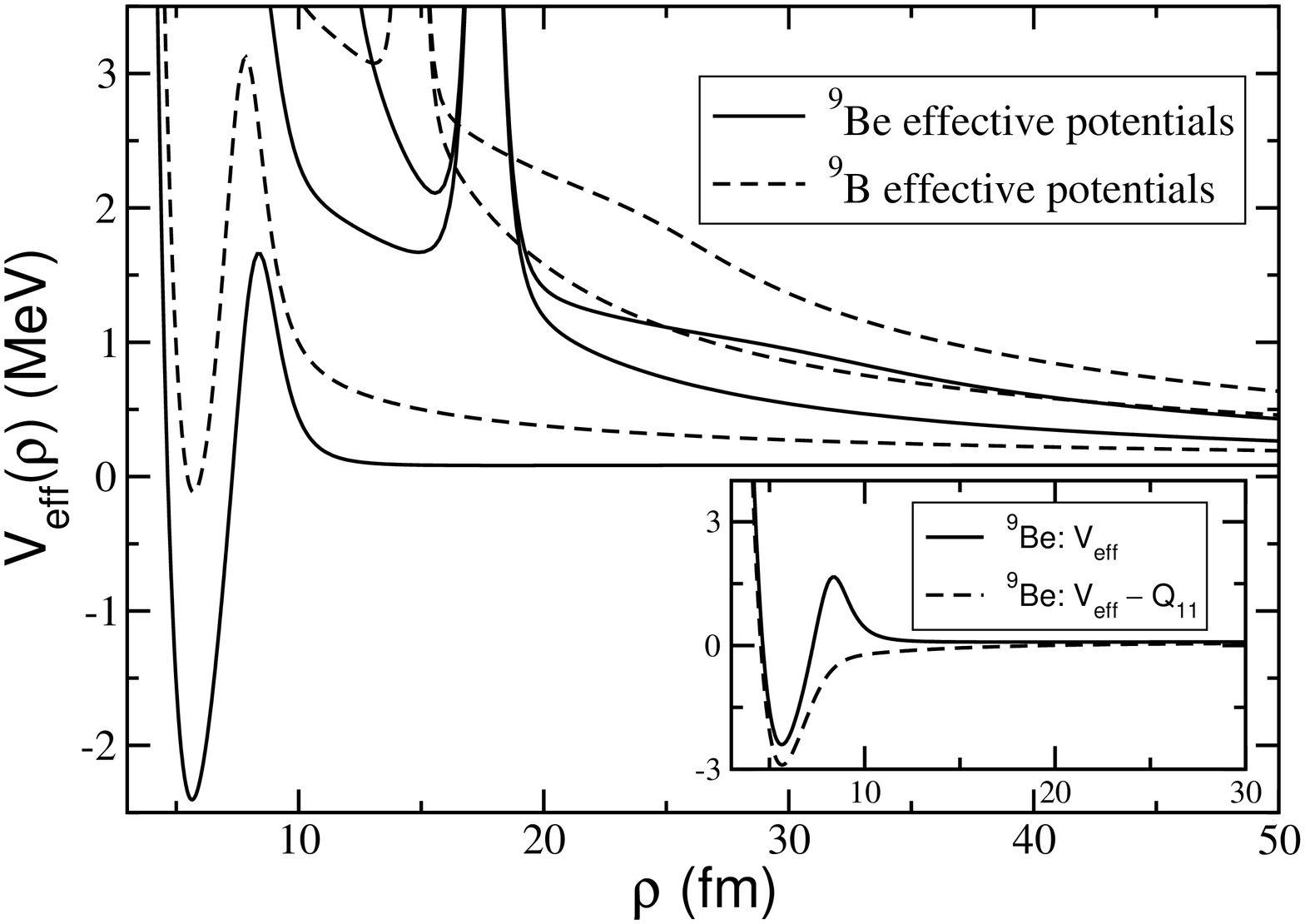}
%/usr/users/asj/raul/Fig1.ps} 
\vspace*{-0.5cm}
\caption{Lowest adiabatic potentials for $^9$Be and $^9$B as a
  function of the hyperradius. The inset shows the $^9$Be lowest
  potential with and without the rearrangement coupling term $Q=
  \langle \phi | \frac{\partial^2}{\partial \rho^2} | \phi
  \rangle_{\Omega}$. }
\label{fig:2}      
%\end{figure}
%\begin{figure}[h]
\vspace*{-0.8cm}
  \includegraphics[width=0.75\textwidth,angle=-90]{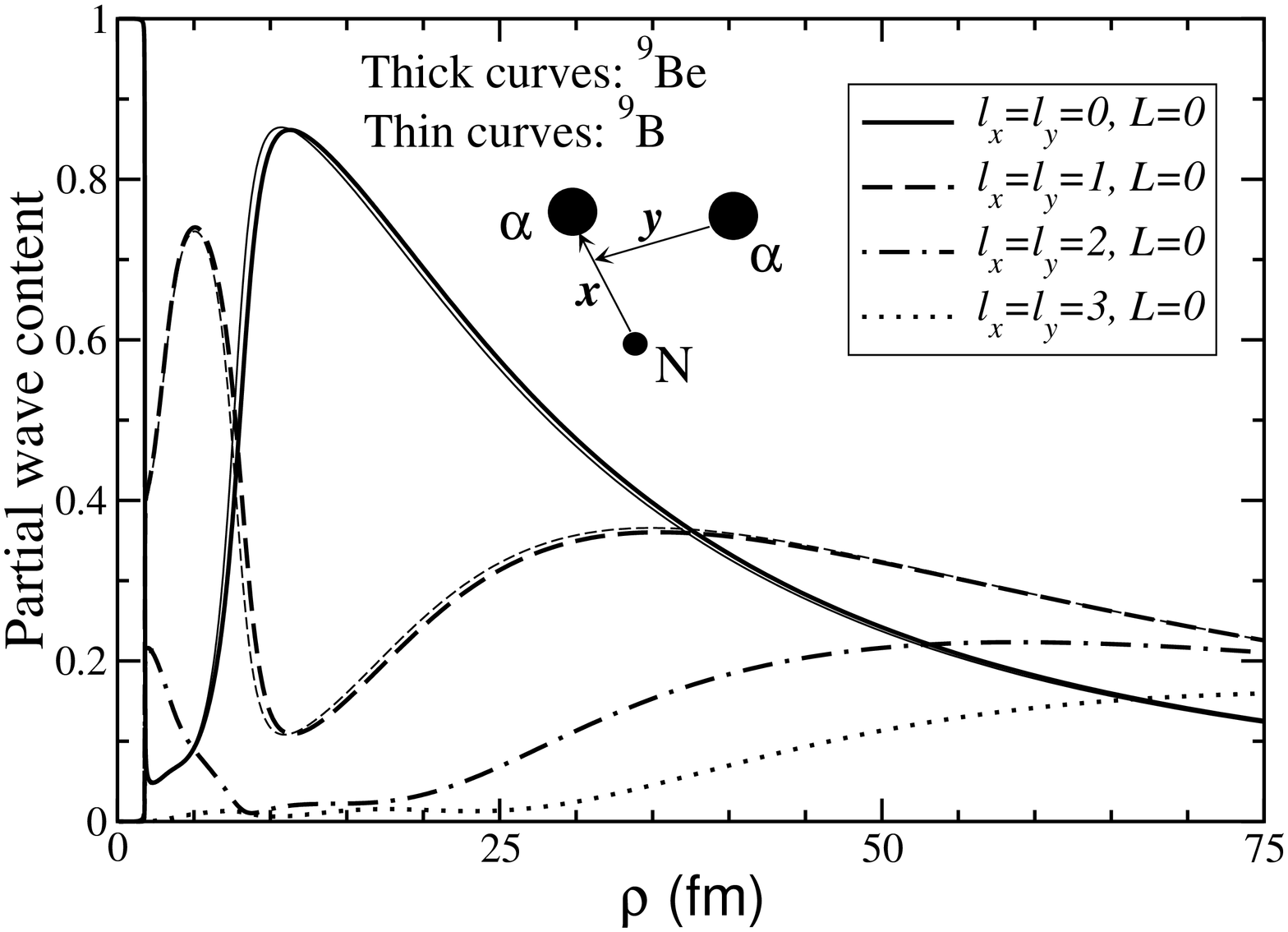}
%/usr/users/asj/raul/Fig2a.ps} 
\vspace*{-0.5cm}
\caption{The partial wave decomposition of the lowest adiabatic 
angular wave function for $^9$Be (thick) and $^9$B (thin) as function
of hyperradius $\rho$. The partial angular momenta $l_x$ and $l_y$
correspond to the coordinates indicated in the figure. For $l_x=l_y=2$
and $l_x=l_y=3$ the curves for $^9$Be and $^9$B can not be distinguished. }
\label{fig:3}      
\end{figure}

The key quantities are the potentials as functions of the hyperradius
as shown in Fig.~\ref{fig:1} where one potential has an attractive
pocket at short distance whereas all the other potentials are
repulsive. The resonance is caught in the pocket and its partial
lifetime, or width, is determined by the tunneling probability through
the barrier.  The coordinate is $\rho$ and {\em a priori} it is not
clear that the width is determined by the barrier of the hyperradial
potential. The angular momentum and parity dependence now follow these
potentials which often lead to non-monotonous behavior. The behavior
at short distances is outside the three-body model and able to
influence the widths substantially in analogy to the effects of
spectroscopic factors \cite{jen10}.

The resonance can change structure from small to large distance
\cite{alv07,alv08}.  In Fig.~\ref{fig:1} the two lowest potentials
cross each other around $\rho$ of 16~fm. The resonance then must
decide whether to maintain its structure determined by the lowest
potential at small distances or gain energy by changing structure. The
compromise can be any continuous division between these extremes.
This dynamic evolution is seen in Fig.~\ref{fig:2} for
$^{9}$Be($1/2^+$) where the inset shows that the barrier is entirely
due to the $Q$-term in Fig.~\ref{e50}, see \cite{gar10}. This implies
that the entire width is due to angular restructuring and in fact
therefore responsible for the resonance character of this state.  This
is demonstrated in Fig.~\ref{fig:3} where the partial wave
decomposition in one Jacobi coordinate set is shown to change
drastically around $\rho$ of 7~fm from a $^5$He$+\alpha$ to a
$^8$Be$+$n structure.  This is consistent with the observation that
the decay products emerge as arising fully from the latter
configuration \cite{sum02}.  In the example of Fig.~\ref{fig:1} this
means that the lowest energy is chosen while the structure is changed
at the level crossing.  These examples show the concepts of dynamic
evolution and decay mechanism \cite{alv10}.

\section{Rates and momentum distributions}
\label{sec:2}

\begin{figure}
\vspace*{+0.3cm}
  \includegraphics[width=0.65\textwidth,angle=-90]{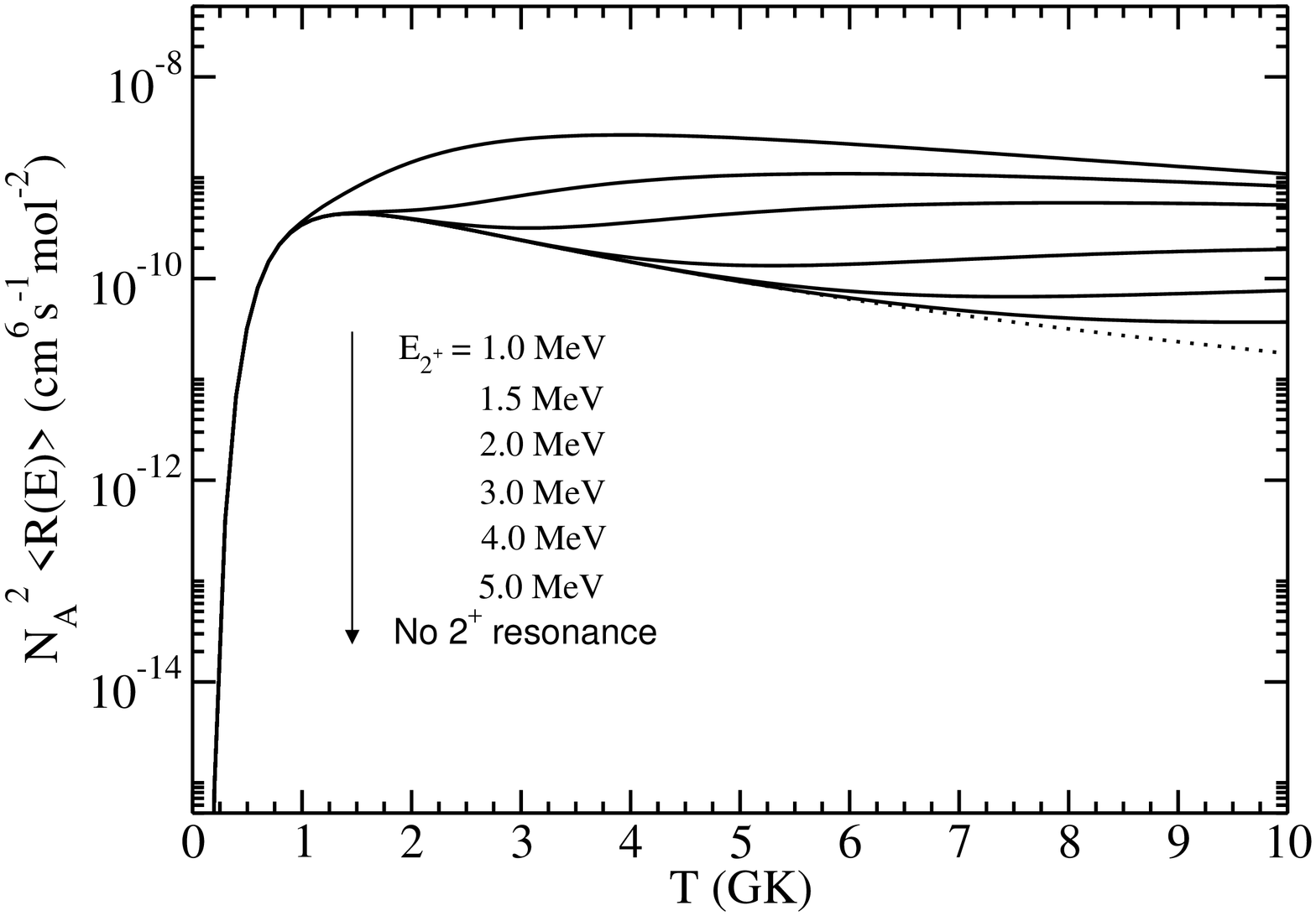}
%/usr/users/asj/raul/Fig3.eps} 
\vspace*{-0.0cm}
\caption{Reaction rate in the sequential case for different energies
  of the lowest 2$^+$ resonance in $^{12}$C. The energy increases from
  the upper curve to the lower from 1 MeV up to 5 MeV. The dotted
  curve is the calculation where the contribution from the
  $2^+\rightarrow 0^+_1$ transition has been completely removed. }
\label{fig:4}      
%\end{figure}
%\begin{figure}[h]
\vspace*{-0.0cm}
  \includegraphics[width=0.65\textwidth,angle=-90]{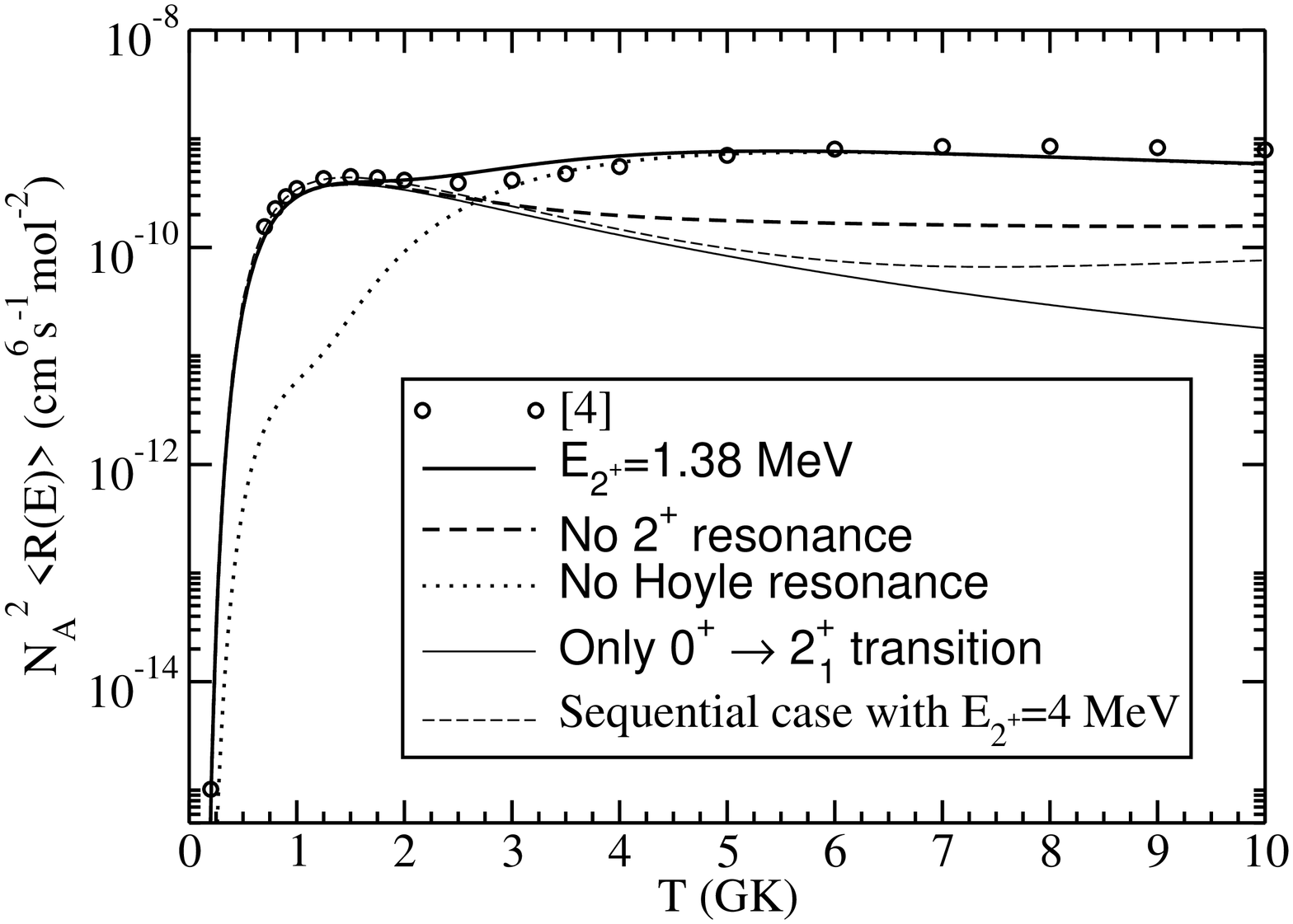}
%/usr/users/asj/raul/Fig4.eps} 
\vspace*{-0.0cm}
\caption{Reaction rate after the full three-body calculation when the
  $2^+$ resonance is placed at 1.38 MeV (thick solid line), when the
  resonance is removed from the calculation (thick dashed line), when
  the full contribution from the $2^+\rightarrow 0^+_1$ transition is
  excluded (thin solid line), and when the Hoyle resonance is removed
  (dotted line). The thin solid curve is the calculation in the
  sequential case when the energy of the $2^+$ resonance is 4.0
  MeV. The open circles are the rate from \cite{ang99}.}
\label{fig:5}      
\end{figure}

The inverse process of three-body decay is the recombination of the
constituent clusters into the bound state of the nucleus for example
by photon emission \cite{die10}. The most prominent as well as most
studied of these processes is the triple $\alpha$ process leading to
$^{12}$C.  At low temperature the lowest $0^+$ resonance, the Hoyle
state, is decisive for the triple $\alpha$ rate which proceeds from
the $0^+$ continuum via an E2-transition to the $2^+$ excited but
bound state.  The process from the $2^+$ continuum is dominated by
E2-transition directly to the $0^+$ ground state.  The position of
the lowest $2^+$ resonance is then important but unknown or at least
controversial.

In Fig.~\ref{fig:4} we show the results of genuine three-body
computations for different positions of the lowest $2^+$ resonance.
The rate varies by almost two orders of magnitude when the energy is
increased from 2~MeV to 5~MeV.  The contribution is only significant
for temperatures above 2-3~GK.  The more realistic uncertainty is seen
in Fig.~\ref{fig:5} where we compare to the standard reference
\cite{ang99} and give the variation from the relatively small $2^+$
energy of 1.38~MeV to complete removal of the $2^+$ resonance. In the
temperature range of around 8~GK the uncertainty in the triple
$\alpha$ rate then comes out to be around a factor of three.

\begin{figure}
  \includegraphics[width=0.57\textwidth,angle=-90]{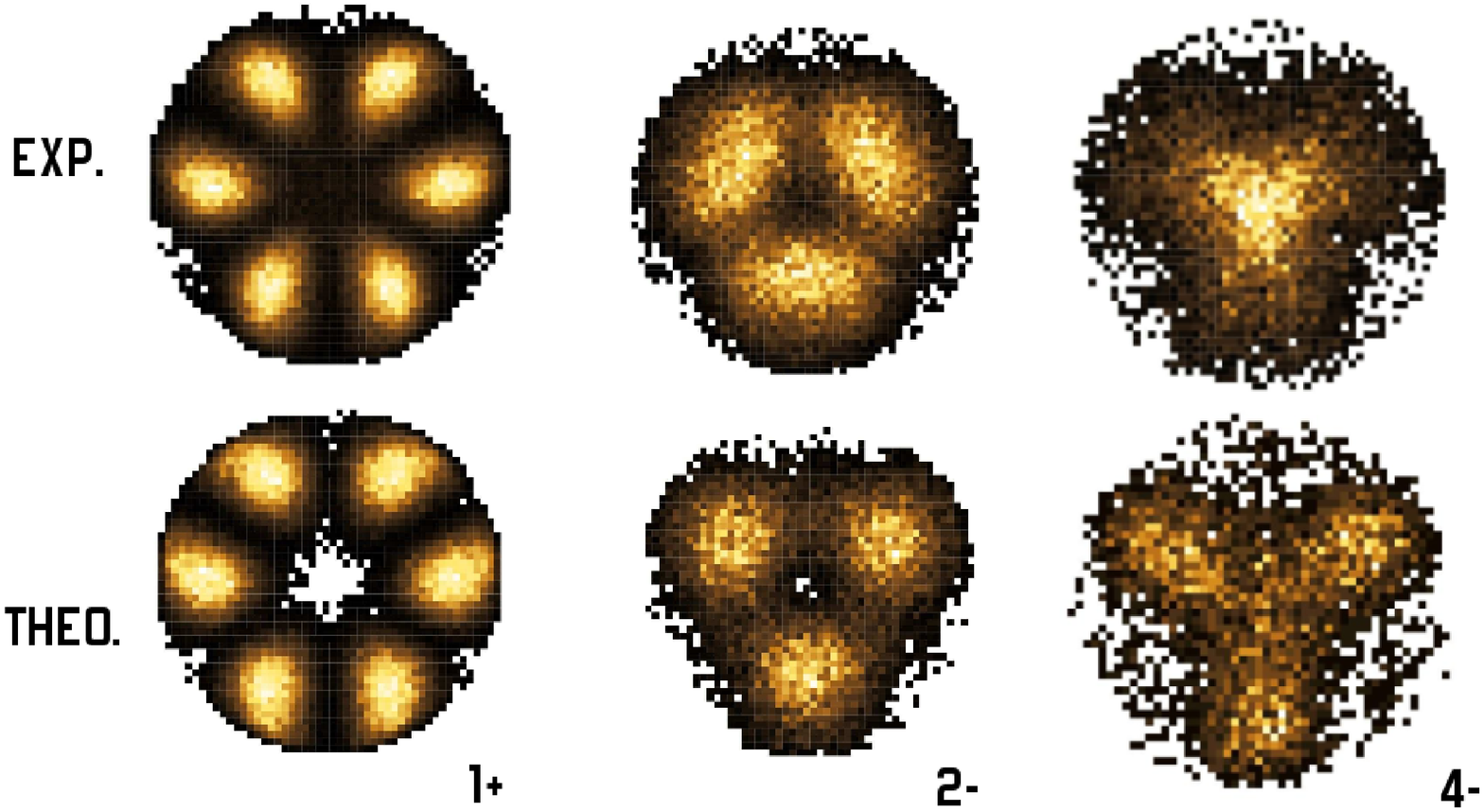}
%/usr/users/asj/raq/th-ex.ps} 
  \caption{The computed Dalitz plots for $^{12}$C $(1^+, 2^-,
    4^-)$-resonances, compared to similar measured
    distributions. O.S. Kirsebom et al. Phys. Rev. C 81, 064313
    (2010).}
\label{fig:6}      
\end{figure}

The large-distance structure of the resonance wave function provides
the momentum distributions of the particles after the three-body decay
\cite{gar07}.  The probability for emission of a particle as function
of its energy is an important part of these distributions. However the
complete information requires two energies for a given total
three-body energy. This gives rise to Dalitz plots which are
two-dimensional probabilities as functions of two independent energies
of the three-body system.  For $^{12}$C decaying into three
$\alpha$-particles the choice could be energies of two different
$\alpha$-particles for a given total resonance energy. We show
computed examples in Fig.~\ref{fig:6} for three resonances compared to
experimental distributions.  The similarities are striking where
especially the zero points in the distributions are interesting. Some
of these are unavoidable as inherent from the angular momentum, parity
and symmetry of the wave function. Other zero points reflect the
dynamical evolution and the decay mechanism and as such they are
significant \cite{fyn09}.

\section{Recombination processes}
\label{sec:3}

\begin{figure}
  \includegraphics[width=0.70\textwidth,angle=-90]{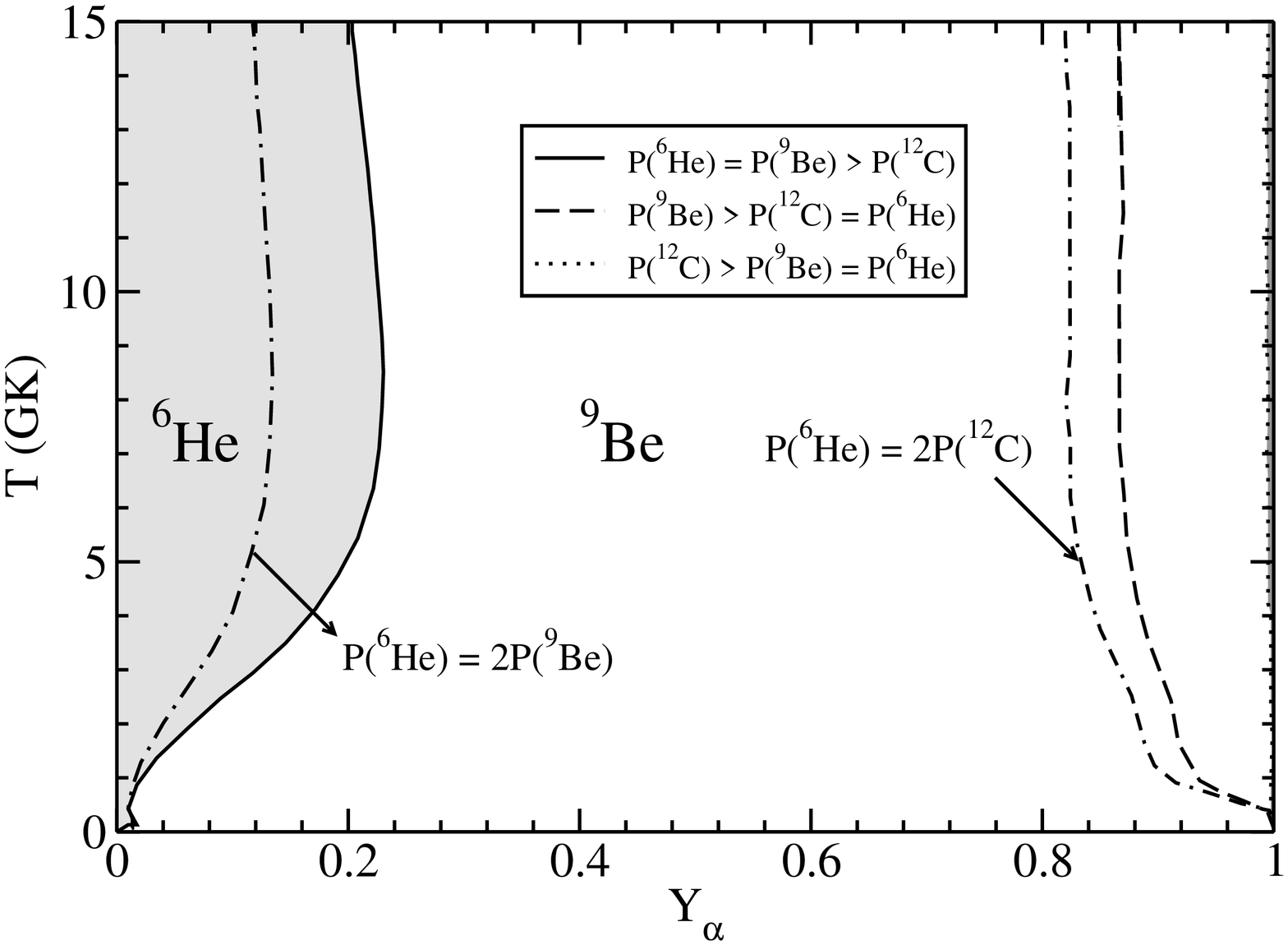}
%/usr/users/asj/raul/Nfig5.eps} 
  \caption{The phase diagram for producing $^6$He, $^9$Be and $^{12}$C
    in the $Y_{\alpha}$-temperature parameter space. The curves
    correspond to a constant ratio of production rates of two nuclei.}
\label{fig:7}      
%\end{figure}
%\begin{figure}
\vspace*{-0.5cm}
  \includegraphics[width=0.75\textwidth,angle=-90]{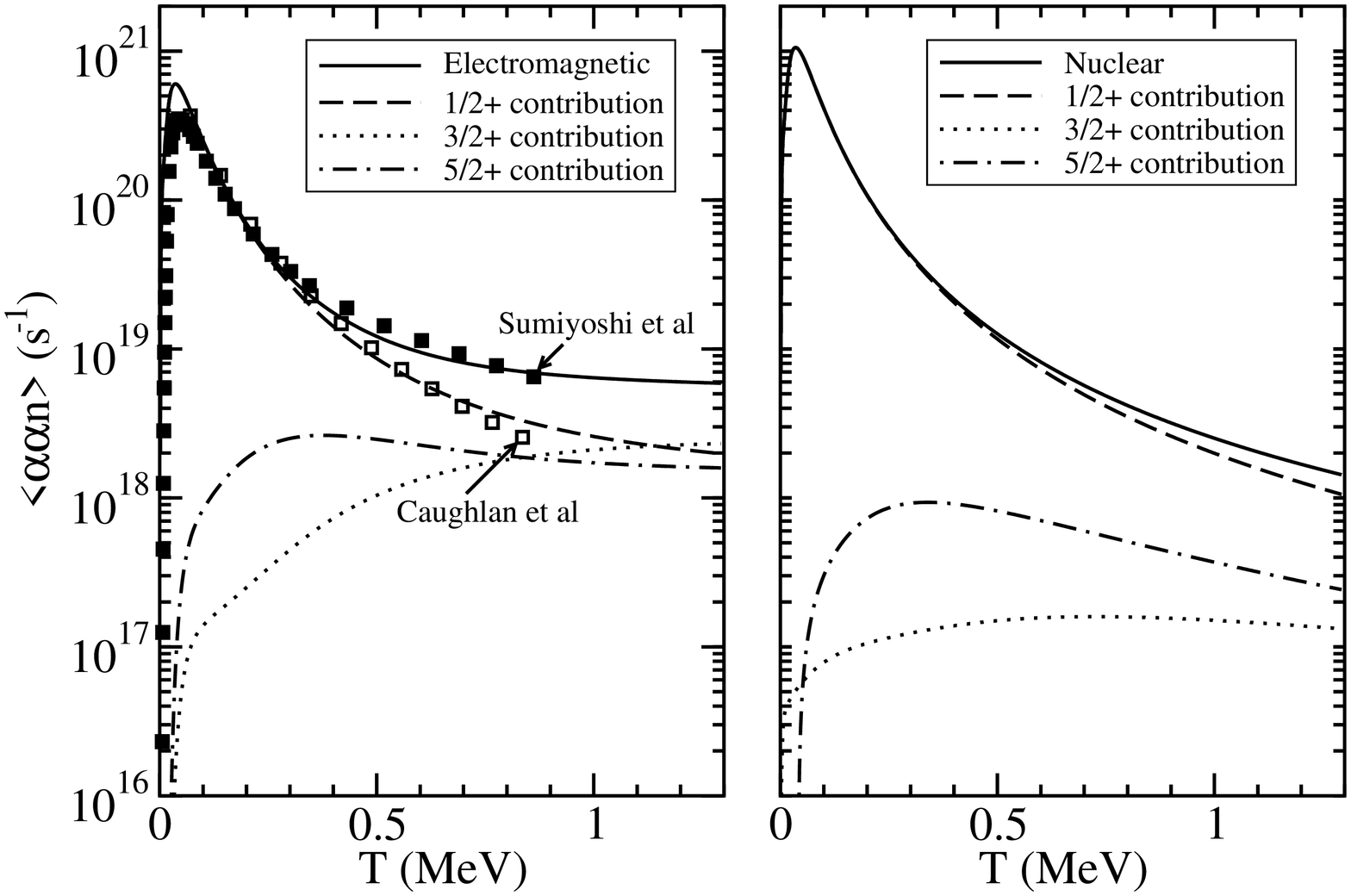}
%/usr/users/asj/raul/fig2m-2.ps} 
\vspace*{-0.5cm}
\caption{The temperature averaged rates for $^9$Be recombination from
  electromagnetic and nuclear processes. The nuclear rate is
  calculated for the neutron density of $10^{30}$~cm$^{-3}$. Sumiyoshi
  et al.: \cite{sum02}, Caughlan et al.: At. Data Nucl. Data Tables
  40, 283 (1988).}
\label{fig:8}      
\end{figure}

The triple $\alpha$ process takes place when $\alpha$-particles are
present in a given volume.  If a mixture of neutrons and
$\alpha$-particles is present two other recombination processes can
also take place, that is creating $^6$He and $^9$Be consisting of one
(two) $\alpha$'s and two (one) neutrons. The relative creation rates
are important for the continuation of the nuclear synthesis leading to
different C-isotopes which in turn are the starting points for nuclear
synthesis into heavier nuclei \cite{die10}.  The density dependence of
these rates is proportional to the number of particles involved in the
given process where two (three) identical components should be counted
twice (thrice).  The temperature dependence is more complicated,
depending for example on resonances, as seen in Figs.~\ref{fig:4} and
\ref{fig:5}.

The density-temperature diagram is shown in Fig.~\ref{fig:7} where
$Y_{\alpha}=N_{\alpha}/(N_{\alpha}+ N_n)$ is the relative fraction of
$\alpha$-particles. Very crudely, $^{12}$C is predominantly created
when the neutron density is low, correspondingly $^6$He is
predominantly created when the neutron density is high, and at
intermediate densities the $^9$Be creation dominates. The details of
when and by how much the different processes contribute is not
obvious.  For example at low $\alpha$-density and low temperature the
$^9$Be rate is larger than that of $^6$He.  This can be traced back to
the low-lying $1/2^+$ resonance in $^9$Be.  These results emerge from
an interplay between recombination rates from individual three-body
continuum states of different angular momentum and parity.

It is well known that radiative capture is much less efficient than
capture where the photon is replaced by a particle of finite mass. In
the environment, where neutrons and $\alpha$'s are present with
certain densities, there is always a finite probability of finding a
fourth particle to substitute the photon and ensure energy and
momentum conservation. This four-body recombination process is largest
when the fourth particle is a neutron because it is neutral (not
pushed away by the Coulomb repulsion) and its $s$-wave interaction
with other neutrons is relatively large \cite{fed10}.  We compare
nuclear and electromagnetic recombination rates in Fig.~\ref{fig:8}
for creating $^9$Be as function of temperature for the continuum
states of different angular momentum.  They are remarkable similar but
this is because we used a rather high neutron density to get
contributions of the same order. The nuclear process has one
additional neutron density as factor compared to the electromagnetic
process. Thus at some density the processes must be of similar size as
shown in Fig.~\ref{fig:8}.

\section{Summary remarks}

We use the hyperspherical adiabatic expansion method to investigate
three-body resonance structures, decay mechanisms, and recombination
rates for selected systems of light nuclei in stellar environments.
We briefly discuss basic ingredients within the method, that is
effective potentials, three-body resonance structure, partial decay
widths, and momentum distributions of particles after the decay of the
resonance. We also investigate recombination of three nuclear clusters
into bound states of a light nuclei. We illustrate by examples of
structure and decay properties of selected resonances in $^6$He,
$^9$Be and $^{12}$C.  Specifically we show results for the influence
of the $2^+$ resonances in $^{12}$C on the triple $\alpha$-rate. We
show the temperature and density dependence of the recombination rates
from neutrons and $\alpha$'s into $^6$He, $^9$Be and $^{12}$C. We
suggest an alternative route to bypass the $A=5,8$ gaps via nuclear
four-body recombination processes.  The possible comparison to
measurements is very favorable.

%\subsection{Subsection title}
%\label{sec:2}
%as required. Don't forget to give each section
%and subsection a unique label (see Sect.~\ref{sec:1}).

%\paragraph{Paragraph headings} Use paragraph headings as needed.

%\begin{figure*}
%  \includegraphics[width=0.75\textwidth]{example.eps}
%\end{figure*}
%

%\begin{table}
%\begin{tabular}{lll}
%\hline\noalign{\smallskip}
%\end{tabular}
%\end{table}

%\begin{acknowledgements}
%\end{acknowledgements}



\end{document}